\documentclass[11pt]{article}
\title{Algebra for generalised quantum observables}
\author{Michael J. W. Hall\\
Theoretical Physics, IAS\\Australian National University\\
Canberra ACT 0200, Australia\\email: michael.hall@anu.edu.au}
\date{}
\begin{document}
\maketitle

\begin{abstract}
Generalised observables (POM observables) are necessary for
representing {\it all} possible measurements on a quantum system. 
Useful algebraic operations such as addition and multiplication are defined 
for these observables, recovering many
advantages of the more restrictive Hermitian
operator formalism.
Examples include new uncertainty relations and metrics, and optical
phase applications.
\end{abstract}

\section{Introduction}

The assertion that {\it all} observable quantities of a quantum system
can be represented by Hermitian operators acting on the Hilbert space of
the system is now well known to be inconsistent with the usual
representation of measurement via interaction between the system and an
apparatus.  In particular, consider the measurement procedure 
defined by preparing
an apparatus in some fixed initial state, allowing an interaction
between the system and apparatus, and measuring some Hermitian operator
on the apparatus.  The probability of a given measurement result, $a$,
for initial state $|\psi\rangle$ of the system, is then found to have the
general form
\begin{equation}  \label{prob}
p(a|\psi) = \langle\psi|A_a|\psi\rangle  ,   
\end{equation}
where $A_a$ is a positive operator acting on the system Hilbert space,
determined by the specific measurement setup (i.e., by the initial
apparatus state, the interaction Hamiltonian, and the apparatus
observable).  

The set of operators $\{A_a\}$ corresponding to such a measurement
procedure is called a probability operator measure (POM) or a positive
operator valued measure (POVM), and the corresponding physical
observable of the system, ${\cal A}$, may be referred to as a POM observable
\cite{helstrom,holevo,kraus,busch}.  
In general $A_a$ is {\it not} a projection operator
associated with an eigenspace of some corresponding Hermitian operator.

Thus, to consider {\it all} possible measurements on a quantum system
(without reference to specific experimental setups), one must consider
all POM observables on the Hilbert space of the system.  Such
observables represent a non-trivial extension of Hermitian observables,
being necessary  to describe, for example,  optimal measurements for
distinguishing between non-orthogonal states \cite{helstrom, holevo,busch,
nonorthog}, and for
describing optical phase {\cite{helstrom, holevo,busch,phase1,phase2,shapiro,
phase3}
and optical heterodyne detection \cite{busch,hetdet1,hetdet2}.

However, what is gained in generality is lost in algebraic simplicity.
For example, it is not {\it a priori} clear how to add and subtract,
much less multiply, POM observables. It 
would therefore be desirable to be able to define an algebra on the
class of POM observables, to regain some of the advantages of the operator
algebra associated with (the less general) Hermitian observables.
The aim of this Letter is to demonstrate the existence of such an
algebra, and applications thereof.

In the following section POM observables are briefly reviewed, and
maximal and non-redundant observables defined.  
In section 3 the sum, product and other binary operations are defined
for the relatively simple 
case of one POM observable and one Hermitian observable, and examples of
statistical deviation and distance given in section 4.  
In section 5 the 
algebraic combination of two arbitrary 
POM observables is considered.  This general
case is more difficult, as certain consistency conditions must be
observed, but is solvable. A generalised uncertainty relation and 
optical phase examples are given in section 6, and
conclusions in section 7.

\section{Generalised observables}

The requirement that the probability distribution $p(a|\psi)$ in
Eq.~(\ref{prob}) is positive and normalised for all states $\psi$
implies that the POM $\{A_a\}$ satisfies
\begin{equation} \label{pom}
A_a\geq 0,~~~~~\sum_a A_a = 1 .
\end{equation}
In fact, any set of operators $\{A_a\}$ satisfying these conditions may be
realised by a measurement procedure of the type discussed in the
Introduction, and hence Eq.~(\ref{pom}) fully characterises the class of
POM observables \cite{helstrom,holevo,kraus,busch}.  The summation is
replaced by integration over continuous ranges of measurement
outcomes.

From Eq.~(\ref{prob}) the expectation value of any function $f({\cal A})$ of
a general POM observable ${\cal A}$ is given by 
\begin{equation} \label{fa}
\langle\, f({\cal A})\,\rangle = \sum_a f(a)\,p(a|\psi) =
\langle\psi |\overline{f({\cal A})}|\psi\rangle  ,
\end{equation}  
where $\overline{f({\cal A})}$ is defined to be the operator
\begin{equation} \label{faop}
\overline{f({\cal A})} := \sum_a f(a) A_a  .
\end{equation}
{\it Hermitian} observables correspond to the special case $A_aA_{a'}=0$ for
$a\neq a'$ (and real-valued outcomes). For such observables the
associated operator $A=\overline{{\cal A}}$ is Hermitian, and satisfies
$f(A) = \overline{f({\cal A})}$ for any function $f$.  Further, one
has $A_a-A_a^2 = A_a\sum_{a'\neq a}A_{a'} =0$, i.e., for Hermitian
POMs each $A_a$ is a projection operator (onto an eigenspace  
associated with eigenvalue $a$ of $A$).

Now, for any positive operator $A_a$ appearing in some POM, a
new POM can be obtained by replacing $A_a$ by 
$N$ copies of $A_a/N$ (i.e., by $A_{a,1},\dots ,A_{a,N}$, with
$A_{a,i}\equiv A_a/N$).  However, this new POM observable is
trivially measured, by making a measurement of the original
observable 
and throwing an $N$-sided die if outcome $a$ is obtained (to distinguish
the $N$ new possibilities).
Thus this new POM is {\it physically redundant}, only
differing from the original in a trivial statistical sense that is
independent of the actual system.  A similar 
redundancy is obtained if the weightings 
$1/N$ are replaced by any set of positive numbers $w_1,\dots ,w_N$ which sum to
unity.  

A POM observable ${\cal A}$ is therefore called {\it non-redundant} if $A_a\neq\lambda
A_b$ for $a\neq b$ and any real number $\lambda$.  
Note that any POM
observable may be trivially (and uniquely) reduced to a non-redundant
POM observable, by adding together any proportional terms in the
corresponding POM.  Note also that
non-redundant observables exclude the trivial possibility $A_a=0$.

Finally, a POM observable ${\cal A}$ is called {\it maximal} if $A_a=|a\rangle\langle
a|$ for all $a$  (the kets $\{|a\rangle\}$ are not necessarily
orthogonal  nor normalised).  It may be shown that
maximal POMs are maximally informative, in the sense that any
measurement
which optimises information gain under a given constraint is equivalent
to the measurement of some maximal POM.
Further, since any positive operator $A_a$ may be
decomposed into a sum of ``back-to-back'' kets, it follows that
every POM observable has a (non-unique) maximal extension.
Attention will therefore be restricted to non-redundant maximal POM
observables in what follows.

\section{Combinations of POM and Hermitian observables}

Defining the algebraic combination of a 
maximal POM observable with any Hermitian
observable is relatively straightforward, and hence worth examining
separately from the general case. The key here is the promotion of the
POM observable to an equivalent  
Hermitian operator on an extended Hilbert space, in
such a way that all algebraic relations between existing Hermitian
operators are preserved. 

In particular, let $H$ denote the Hilbert space of the system, and 
for a maximal POM observable ${\cal A}\equiv\{|a\rangle\langle a|\}$
let $H_{\cal A}$ denote the Hilbert space of square-integrable functions
over the space of outcomes of ${\cal A}$.  There is then an orthonormal basis
$\{|a)\}$ for $H_A$ associated with these outcomes, and 
a natural mapping from $H$ to $H_A$ defined by 
\begin{equation} \label{psiext}
|\psi\rangle\rightarrow|\psi_{\cal A}) := \sum_a \langle a|\psi\rangle \, |a)  .
\end{equation}
From Eqs.~(\ref{prob}) and (\ref{psiext}) one has 
\[ p(a|\psi) = |\langle a|\psi\rangle|^2 = |(a|\psi_{\cal A})|^2  . \]
Thus the POM $\{|a\rangle\langle a|\}$ on $H$ has statistics 
equivalent to the  Hermitian operator
\begin{equation} \label{aext}
\hat{A} := \sum_a a\, |a)(a|  
\end{equation}
on $H_{\cal A}$, where the latter corresponds to the {\it Hermitian} POM
$\{|a)(a|\}$.

Note that the representation $\psi(a)=\langle a|\psi\rangle =
(a|\psi_{\cal A})$ is well known for particular POM observables (eg, the
coherent state and canonical phase representations \cite{holevo,shapiro}).  The
associated representation of ${\cal A}$ as a Hermitian operator $\hat{A}$ on $H_A$ 
provides a simple example of Naimark's extension theorem
\cite{helstrom, holevo,busch} for the (admittedly trivial) case of maximal POMs.

The set of extended states, $\{|\psi_{\cal A})\}$, is characterised by
the projection operator
\begin{equation} \label{e}
E := \sum_{a,a'} \langle a|a'\rangle |a)(a'| = E^2 .
\end{equation}  
In particular, it may be checked that any normalised ket $|\psi)$ in
$H_{\cal A}$ is generated by some
physical state $|\psi\rangle$, as per Eq.~(\ref{psiext}), if and only if
\[  E|\psi) = |\psi)  . \]
Thus the unit eigenspace of $E$ is isomorphic to the physical Hilbert
space $H$.  Note that for a {\it Hermitian} observable one has $E\equiv
1$, and hence in this case (and only in this case) $H_{\cal A}$ is isomorphic
to $H$.

Further, let $X$ be any Hermitian operator on $H$.  Then there is a
natural extension of $X$ to a Hermitian operator $X_{\cal A}$
on $H_{\cal A}$, defined by
\begin{equation} \label{xext}
X_{\cal A} = \sum_{a,a'} \langle a |X|a'\rangle\,|a)(a'|  .
\end{equation}
It may be checked that (i) the product $XY$ is mapped to 
$X_{\cal A}Y_{\cal A}$, and (ii) the 
state $X|\psi\rangle$ is
mapped to $X_{\cal A}|\psi_{\cal A})$.  Thus this extension
preserves all algebraic properties of Hermitian operators
and states on $H$.

Finally, there is an inverse mapping of states and 
observables from $H_{\cal A}$ to $H$,
generated by the projection $|a)\rightarrow |a\rangle$.
It follows from Eqs.~(\ref{psiext}), (\ref{aext}) 
and (\ref{xext})
that $|\psi_{\cal A})\rightarrow |\psi\rangle$ and 
$X_{\cal A}\rightarrow 
X$ under this
mapping, and that the Hermitian POM $\{|a)(a|\}$ corresponding
to $\hat{A}$ is mapped to the original POM $\{|a\rangle\langle
a|\}$. Moreover, an arbitrary POM $\{|k)(k|\}$ on $H_{\cal A}$ is
mapped to the POM observable $
\{|{k}\rangle\langle{k}|\}$ on $H$, where
\begin{equation} \label{projk}
|{k}\rangle := \sum_a (a|k) |a\rangle  .
\end{equation}
Note that orthonormality of the basis $\{|a)\}$ implies 
\begin{eqnarray*}
\sum_{{k}} |{k}\rangle\langle{k}| & = &
\sum_{k,a,a'} (a|k)(k|a')|a\rangle\langle a'| \\ & = &
\sum_{a,a'} (a|a')|a\rangle\langle a'| \\ & = &
\sum_a |a\rangle\langle a| = 1  ,
\end{eqnarray*}
as required by Eq.~(\ref{pom}).

The tools for defining algebraic combinations of ${\cal A}$ and $X$ (such as
the sum ${\cal A}+X$, the symmetric product $({\cal A}X+X{\cal A})/2$, and the ``commutator''
$i[{\cal A},X]$), are now laid out.  In particular, 
let $g$ denote {\it any} function which maps pairs of Hermitian operators to a
Hermitian operator.  Hence,
\begin{equation} \label{kdef}
\hat{K} := g(\hat{A},X_{\cal A}) 
\end{equation}
is a Hermitian operator on $H_{\cal A}$, with an associated Hermitian POM
$\{|k)(k|\}$.  The corresponding mapping $g({\cal A},X)$, for a general POM
observable ${\cal A}$ and Hermitian observable $X$ on H, is then defined to be
the POM observable
\begin{equation} \label{gax} 
g({\cal A},X):={K}\equiv\{|{k}\rangle\langle{k}|\} ,
\end{equation}
where $|{k}\rangle$ is defined by Eq.~(\ref{projk}). 

Note from Eqs.~(\ref{psiext}) and (\ref{projk}) that
$\langle{k}|\psi\rangle = (k|\psi_{\cal A})$. Hence the
probability distributions $p(k|\psi)$ and $p(k|\psi_{\cal A})$ are
equivalent, and from Eqs.~(\ref{kdef}) and (\ref{gax}) one therefore 
has the identity
\begin{equation} \label{expec}
\langle \,g({\cal A},X)\,\rangle = (\psi_{\cal A}|g(\hat{A},X_{\cal A})|\psi_{\cal A}) = \sum_{a,a'}\langle\psi|a\rangle\, (a|g(\hat{A},X_{\cal A})|a')\,\langle a'|\psi\rangle
\end{equation}
for expectation values.  This
is a very useful formula, as it means one can calculate the
expectation value of $g({\cal A},X)$ without having to explicitly determine the
corresponding POM $\{{k}\rangle\langle{k}|\}$ (which would 
require diagonalising the Hermitian operator
$\hat{K}=g(\hat{A},X_{\cal A})$). 

\section{Examples: deviation and distance}

From Eq.~(\ref{expec}) one may calculate the statistical 
deviation between ${\cal A}$ and $X$,
for a given state $|\psi\rangle$, as
\begin{eqnarray} \label{dev}
\langle \,(X-{\cal A})^2\,\rangle & = & (\psi_{\cal A}|X_{\cal A}^2 + \hat{A}^2 -X_{\cal A}\hat{A}-
\hat{A}X_{\cal A}|\psi_{\cal A})\nonumber
\\& = & \langle X^2\rangle + \langle {\cal A}^2\rangle - 
\sum_{a,a'} \left[ (\psi_{\cal A}|a)(a|X_{\cal A}|a')(a'|\hat{A}|\psi_{\cal A})+ c.c.\right]\nonumber
\\ & = & \langle\psi|(X-\overline{{\cal A}})^2|\psi\rangle +
\langle\psi|\overline{{\cal A}^2} - (\overline{{\cal A}})^2|\psi\rangle , 
\end{eqnarray}
where Eqs.~(\ref{psiext}), (\ref{aext}) and (\ref{xext}) have been used,
and the operators $\overline{{\cal A}}$, $\overline{{\cal A}^2}$ are defined via 
Eq.~(\ref{faop}). This quantity provides a measure of the degree to
which the POM ${\cal A}$ may be approximated by a given Hermitian operator $X$.  
Note that the second term is {\it not} equal to the variance of
${\cal A}$. 
One also has the alternative formula
\begin{eqnarray*}
\langle \,(X-{\cal A})^2\,\rangle & = & \sum_a (\psi_{\cal A}|(X_{\cal A}-\hat{A})|a)(a|
X_{\cal A}-\hat{A}|\psi_{\cal A}) \\ & = & 
\sum_a |\langle a|X-a|\psi\rangle|^2 , 
\end{eqnarray*}
which has been used previously (without algebraic justification) in
the context of obtaining exact uncertainty relations between photon
number and optical phase \cite{eurphase}. 

Further, the ``distance'' between a POM observable ${\cal A}$ and a Hermitian
observable $X$ on $H$ may be defined via 
a natural generalisation of the Hilbert-Schmidt metric:
\begin{eqnarray} 
d({\cal A},X)^2 & := & {\rm tr}[\, (\hat{A}-X_{\cal A})^2\,]
\nonumber\\ \label{metric} & = & {\rm tr}[(X-\overline{{\cal A}})^2 +
\overline{{\cal A}^2}-(\overline{{\cal A}})^2] .
\end{eqnarray}
This generalised metric satisfies the triangle inequalities
\[ d({\cal A},X) + d({\cal A},Y)\geq d(X,Y)\geq |d({\cal A},X) - d({\cal A},Y)| \]
as an automatic consequence of the Hilbert-Schmidt metric on $H_{{\cal A}}$,
and hence $d({\cal A},X)$ may indeed be interpreted as a measure of ``distance''. 

From Eq.~(\ref{metric}) one has the Pythagorean relation
\[ d({\cal A},X)^2 = d({\cal A},\overline{{\cal A}})^2 + d(\overline{{\cal A}},X)^2 \]
between ${\cal A}$, $X$ and $\overline{{\cal A}}$.  It follows immediately that $\overline{{\cal A}}$
represents the Hermitian operator ``closest'' to ${\cal A}$, being separated
from ${\cal A}$ by the 
minimum distance 
\[ \min_X d({\cal A},X) = d({\cal A},\overline{{\cal A}}) = \{{\rm tr}[\overline{{\cal A}^2} -
(\overline{{\cal A}})^2]\}^{1/2}. \]
This minimum distance is a useful measure of the inherent
``fuzziness'' of ${\cal A}$, vanishing if and only if ${\cal A}$ is a Hermitian observable.
 
\section{Combining arbitrary POM observables}

To generalise the above results to algebraic combinations of
two maximal POMs ${\cal A}\equiv\{|a\rangle\langle a|\}$,
${\cal B}\equiv\{|b\rangle\langle b|\}$,  
one must find a Hilbert space $H_{\cal AB}$ which contains two suitable
orthonormal sets $\{|a)\}$, $\{|b)\}$. It turns out that, 
to avoid such undesirable properties such as ${\cal A}-{\cal A}\neq 0$, the compatibility
of ${\cal A}$ and ${\cal B}$ must explicitly be taken into account.  The resulting
simultaneous
mapping of ${\cal A}$ and ${\cal B}$ to Hermitian operators on $H_{{\cal AB}}$ corresponds to a novel and 
highly nontrivial Naimark extension, in
contrast to the mapping from $H$ to $H_{\cal A}$ defined in section 3.
 
First, note that each state $|\psi\rangle$
on $H$ will have two possible extensions on $H_{{\cal AB}}$, given by $|\psi_{\cal A})$ and 
$|\psi_{\cal B})$ as calculated via Eq.~(\ref{psiext}), corresponding to
respective projection operators $E_{\cal A}$ and $E_{\cal B}$ defined as per
Eq.~(\ref{e}). 
Consistency requires that these are equal.
Calculating $(b|\psi_{\cal A})$
and $(a|\psi_{\cal B})$ via Eq.~(\ref{psiext}) then yields the conditions
\begin{equation} \label{psiab}
|a\rangle =\sum_b (b|a)|b\rangle,~~~|b\rangle = \sum_a (a|b)|a\rangle
\end{equation}
on the inner product $(a|b)$.

Second, since the statistics of ${\cal A}$ and ${\cal B}$ on $H$ will be equivalent to those
of the Hermitian operators
$\hat{A}$ and $\hat{B}$ on $H_{{\cal AB}}$ respectively,
as calculated via Eq.~(\ref{aext}), it  
is physically desirable that any {\it identical} statistical components of 
${\cal A}$ and ${\cal A}$ are mapped
to identical components of $\hat{A}$ and $\hat{B}$, i.e., 
\begin{equation} \label{comp}
|a)(a| = |b)(b|~~~{\rm for}~~~|a\rangle\langle a|=|b\rangle\langle b| .
\end{equation}
This condition ensures that compatible aspects of the observables ${\cal A}$ and
${\cal B}$ are preserved by the extension to $H_{{\cal AB}}$ (and in particular that
$\hat{A}=\hat{B}$ for ${\cal A} = {\cal B}$).

If conditions (\ref{psiab}) and (\ref{comp}) can be satisfied, then 
any function $g$ which maps pairs of Hermitian
operators to a Hermitian operator will define a 
Hermitian operator
\[ \hat{K}:= g(\hat{A},\hat{B}) , \]
on $H_{{\cal AB}}$, with an associated Hermitian POM 
$\{|k)(k|\}$.  The corresponding
POM observable $g({\cal A},{\cal B})$ on $H$ is then defined to be the POM ${K} \equiv
\{|{k}\rangle\langle{k}|\}$, with
\[ |{k}\rangle :=\sum_a(a|k)|a\rangle = \sum_b(b|k)|b\rangle \]
in analogy to Eq.~(\ref{projk}), where the second equality follows 
immediately from Eq.~(\ref{psiab}). 

It therefore remains to find orthonormal sets $\{|a)\}$, $\{|b)\}$ 
satisfying Eqs.~(\ref{psiab}) and (\ref{comp}).  
First, let $C:=\{|a\rangle\langle a|\}\cap \{|b\rangle\langle b|\}\equiv\{|c\rangle\langle c|\}$ denote the
set of {\it common} elements of the POMs corresponding to ${\cal A}$ and ${\cal B}$.
$C$ is physically well-defined for non-redundant POMs, where 
measurement of the POM
$C\cup\{ C_0\}$, with
\[ C_0 := 1 - \sum_c |c\rangle\langle c| , \]
corresponds to the simultaneous measurement of the compatible components of 
${\cal A}$ and ${\cal B}$. 
Condition~(\ref{comp}) implies that these compatible components are promoted
to simultaneous eigenstates of $\hat{A}$ and $\hat{B}$.  

Decomposing the POMs for ${\cal A}$ and ${\cal B}$ as $\{|c\rangle\langle c|,
|\tilde{a}\rangle\langle \tilde{a}|\}$ and $\{|c\rangle\langle c|, 
|\tilde{b}\rangle\langle \tilde{b}|\}$ respectively, it follows that
\begin{equation} \label{sumc}
\sum_{\tilde{a}} |\tilde{a}\rangle\langle\tilde{a}| = \sum_{\tilde{b}}
|\tilde{b}\rangle\langle \tilde{b}| = C_0 .
\end{equation}
Eqs.~(\ref{psiab}) and (\ref{comp}) then imply that the orthonormal set
$\{|c)\}$ is orthogonal to each of the orthonormal sets
$\{|\tilde{a})\}$, $\{|\tilde{b})\}$ on $H_{{\cal AB}}$, and further that
the inner product $(\tilde{a}|\tilde{b})$ satisfies
\[
|\tilde{a}\rangle = \sum_{\tilde{b}} (\tilde{b}|\tilde{a}) |\tilde{b}\rangle,
~~~|\tilde{b}\rangle = \sum_{\tilde{a}} (\tilde{a}|\tilde{b})
|\tilde{a}\rangle .  \]
These equations are formally solved by the choice
\begin{equation} \label{ab}
(\tilde{a}|\tilde{b}) = \langle \tilde{a}|C_0^{-1}|\tilde{b}\rangle 
\end{equation}
as may easily be checked using Eq.~(\ref{sumc}).
However, it must be verified that this solution 
has all the properties required of an inner product
between two orthonormal (not necessarily complete) sets in some Hilbert
space $\tilde{H}_{{\cal AB}}$ (one may then take
$H_{{\cal AB}}=\tilde{H}_{{\cal AB}}\oplus H_C$, where $H_C$ is the span of $\{|c)\}$).

First, each of $|\tilde{a}\rangle$ and $|\tilde{b}\rangle$ are
orthogonal to 
the zero-eigenspace of $C_0$ (since Eq.~(\ref{sumc}) implies that
$\langle \tilde{a}|C_0|\tilde{a}\rangle$ and $\langle
\tilde{b}|C_0|\tilde{b}\rangle$ are strictly positive), and hence 
the righthand side of Eq.~(\ref{ab}) is well-defined.    
Second, noting $C_0$ is Hermitian, then 
$(\tilde{a}|\tilde{b})=(\tilde{b}|\tilde{a})^*$ as required.
Finally, for it to be possible to write 
each $|\tilde{a})$ as an orthogonal superposition of the
$|\tilde{b})$ and some orthonormal set $\{|x)\}$, i.e.,
\[ |\tilde{a}) = \sum_{\tilde{b}} (\tilde{b}|\tilde{a}) |\tilde{b}) +
\sum_{x} (x|\tilde{a})|x),  \]
one must have $\sum_{\tilde{b}}|(\tilde{b}|\tilde{a})|^2\leq 1$ (and
similarly $\sum_{\tilde{a}}|(\tilde{b}|\tilde{a})|^2\leq 1$).  But
\[ \sum_{\tilde{b}}|(\tilde{b}|\tilde{a})|^2 = \langle\tilde{a}|C_0^{-1}
|\tilde{a}\rangle = \langle \overline{a}|\overline{a}\rangle  \]
from Eq.~(\ref{ab}), where
$|\overline{a}\rangle:=C_0^{-1/2}|\tilde{a}\rangle$.  Noting from
Eq.~(\ref{sumc}) that
$\sum_{\overline{a}} |\overline{a}\rangle\langle\overline{a}|$ is
equivalent to the unit operator on the span of $\{|\overline{a}\rangle\}$,
one then has
\begin{eqnarray*}
 \langle \overline{a}|\overline{a}\rangle & = & \langle
\overline{a}|[ \sum_{\overline{a}'}| 
\overline{a}'\rangle\langle\overline{a}'] |\overline{a}\rangle \\
& = & \sum_{\overline{a}'} |\langle \overline{a}|\overline{a}'\rangle|^2
\geq \langle \overline{a}|\overline{a}\rangle^2 , 
\end{eqnarray*} 
implying $\langle \overline{a}|\overline{a}\rangle\leq 1$ as required.

\section{Examples: uncertainty relations and phase}

The expectation value
(and hence the statistics) of any algebraic combination of 
two arbitrary POMs ${\cal A}$ and ${\cal B}$ on $H$ may now be calculated via the relation
\begin{equation} \label{genexpec}
\langle \,g({\cal A},{\cal B})\,\rangle = (\psi|g(\hat{A},\hat{B})|\psi)
\end{equation}
analogous to Eq.~(\ref{expec}), where $|\psi)$ denotes either of
$|\psi_{\cal A})$, $|\psi_{\cal B})$. 

The examples of statistical deviation and distance in section 4
generalise immediately to such pairs of POM observables.
 As a further example, the expectation of the ``commutator'' of ${\cal A}$ and ${\cal B}$
follows as
\begin{eqnarray*}
\langle\,i[{\cal A},{\cal B}]\,\rangle & = & i(\psi|[\hat{A},\hat{B}]|\psi)\\
& = & i(\psi|[\sum_{a,b} ab(a|b)|a)(b| - h.c. ]|\psi)\\
& = & i\langle\psi|[\sum_{\tilde{a},\tilde{b}}
\tilde{a}\tilde{b}\langle\tilde{a}|C_0^{-1}|\tilde{b}\rangle
|\tilde{a}\rangle\langle\tilde{b}| -h.c. ]|\psi\rangle\\
& = & i\langle\psi|\tilde{A}C_0^{-1}\tilde{B} -
\tilde{B}C_0^{-1}\tilde{A}|\psi\rangle ,
\end{eqnarray*}
where  the operators 
\[ \tilde{A}:=\sum_{\tilde{a}} \tilde{a}|\tilde{a}\rangle
\langle\tilde{a}|,~~~\tilde{B}:=\sum_{\tilde{b}}
\tilde{b}|\tilde{b}\rangle\langle\tilde{b}|  \]
represent the restrictions of $\overline{{\cal A}}$
and $\overline{{\cal B}}$ to non-identical $|a\rangle\langle a|$ and
$|b\rangle\langle b|$. 
The Heisenberg uncertainty relation on $H_{{\cal AB}}$ therefore leads 
immediately to the {\it generalised} uncertainty relation
\begin{equation}
\Delta {\cal A}\,\Delta {\cal B} \geq \frac{1}{2}|\langle\psi|\tilde{A}C_0^{-1}\tilde{B}
- \tilde{B}C_0^{-1}\tilde{A}|\psi\rangle|
\end{equation}
for arbitrary POM observables ${\cal A}$ and ${\cal B}$ on $H$.

In the case that ${\cal A}$ and ${\cal B}$ have {\it no} POM components in common, then $C_0=1$ and the above uncertainty relation reduces to
\[  \Delta {\cal A}\,\Delta {\cal B} \geq
\frac{1}{2}|\langle\psi|[\overline{{\cal A}},\overline{{\cal B}}|\psi\rangle| . \]
As an example, for the photon number operator
$N=\sum_nn|n\rangle\langle n|$ and canonical phase POM
$\Phi\equiv\{|\phi\rangle\langle\phi|\}$, where
$|\phi\rangle=(2\pi)^{-1/2}\sum_ne^{in\phi}|n\rangle$ and  
$\phi\in(-\pi,\pi]$ \cite{helstrom,holevo,busch,phase1,phase2,shapiro,phase3}, 
a straightforward calculation yields 
\begin{eqnarray*}
\Delta N\Delta\Phi & \geq & (1/2)|\langle\psi|[N,\overline{\Phi}]|\psi\rangle|
\\& = & (1/2)|\langle\psi|\sum_{m,n}(-1)^{m+n}|m\rangle\langle n|
 -1|\psi\rangle|
\\& = & (1/2)|1-2\pi p(\pi|\psi)| ,
\end{eqnarray*}
in agreement with previous (non-algebraic) methods 
\cite{shapiro,eurphase,pegg}.

Another phase observable of interest arises from ideal heterodyne
detection, where one makes a simultaneous (but noisy) measurement of the
quadratures of a single-mode optical field \cite{busch,hetdet1,hetdet2}.  This
measurement is represented by the
the so-called ``coherent-state'' POM
${\cal A}_H\equiv\{\pi^{-1}|\alpha\rangle\langle\alpha|\}$, where $|\alpha\rangle$
denotes the coherent state corresponding to eigenvalue 
$\alpha=re^{i\phi}$ of the
photon annihilation operator $a$.
Thus $\Phi_H:=\arg {\cal A}_H$ is a corresponding phase
observable: the {\it heterodyne} phase \cite{busch,hetphase,hallfuss} (see also
Ref.~\cite{paul} for a different realisation of this observable).

Since phase is a periodic variable, the {\it circular} deviation 
\[ \delta_H := 1-|\langle \,e^{i\Phi}e^{-i\Phi_H}\,\rangle |   \] 
provides a more natural measure of comparison for $\Phi$ and $\Phi_H$ than 
does
$\langle(\Phi-\Phi_H)^2\rangle$ (even better is $1-|\langle e^{i(\Phi-\Phi_H)}
\rangle|$, but this is more difficult to evaluate). 
This quantity will be close to zero in cases where heterodyne phase
provides a good approximation to the canonical phase, and close to unity 
in cases where heterodyne phase is a poor estimate of the canonical
phase.  Since $\delta_H$ is defined by 
an algebraic function of two POM observables, it can be calculated by
the above methods.

In particular, from Eq.~(\ref{genexpec}) one has $\langle
\,{\cal AB}\,\rangle = \langle\psi|\overline{{\cal A}}\,\,\overline{{\cal B}}|\psi\rangle$ 
when ${\cal A}$ and ${\cal A}$ share no common components. 
A straightforward calculation gives
\[ \overline{e^{i\Phi}} = \sum_n |n\rangle\langle n+1| , \]
while a Gaussian integration yields \cite{hallfuss,paul}) 
\begin{eqnarray*}
\overline{e^{-i\Phi_H}} & = & \pi^{-1} \int d^2\alpha\, e^{-i\phi}
|\alpha\rangle\langle\alpha| \\ 
& = & \sum_n \Gamma(n+3/2)(n!)^{-1}(n+1)^{-1/2}|n+1\rangle\langle n| , 
\end{eqnarray*} 
and hence one finds
\begin{equation} \label{delta} 
\delta_H = 1 - \sum_n |\langle n|\psi\rangle|^2
\Gamma(n+3/2)(n!)^{-1}(n+1)^{-1/2} .
\end{equation}

The circular
deviation between the canonical and heterodyne phases is therefore
completely determined by the photon number distribution of the state.
Further, Stirling's formula for the Gamma function may be used to obtain the
asymptotic formula
\[ \delta_H \sim \langle\psi|(N+1)^{-1}|\psi\rangle/8  \]
from Eq.~(\ref{delta}), to first order in $1/(N+1)$. Hence 
$\delta_H$ is typically small for high energy states, implying that   
heterodyne detection provides an accurate estimate of the canonical
phase for such states.

\section{Conclusions}

It has been shown how to define algebraic operations for arbitrary pairs
of generalised quantum observables.  As well as being of intrinsic
interest, this allows many of the advantages of calculations with
Hermitian operators to be realised for the more general POM observables,
as indicated by the examples in sections 4 and 6.  

The algebraic combination of any two POM
observables ${\cal A}$ and ${\cal B}$ with any number of Hermitian observables
$X,Y,Z,\dots$ may also be defined, with Eqs.~(\ref{expec}) and
(\ref{genexpec}) generalising to 
\[ \langle\, g({\cal A},{\cal B},X,Y,Z,\dots)\,\rangle  =
(\psi_{{\cal AB}}|g(\hat{A},\hat{B},X_{{\cal AB}},Y_{{\cal AB}}, Z_{{\cal AB}},\dots)|\psi_{{\cal AB}}) , \]
where $|\psi_{{\cal AB}})$ denotes either of $|\psi_{\cal A})\equiv|\psi_{{\cal B}})$ and
$X_{{\cal AB}}$ denotes either of $X_{\cal A}\equiv X_{\cal B}$ (the second equivalence
follows from the first, via $E_{\cal A}\equiv E_{\cal B}$).  Note that for the special case where ${\cal A}$ and ${\cal B}$ correspond to two Hermitian operators $A$ and $B$ respectively, then $H_{\cal AB}$ is isomorphic to $H$, and so $g({\cal A},{\cal B},X,Y,Z,\dots)$ becomes equivalent to the Hermitian operator $g({A},{B},X,Y,Z,\dots)$.

A number of issues remain for future investigation.  First, only binary
combinations of POM observables have been considered.  It is not clear
whether general combinations of three or more such observables can be
consistently defined, nor even whether, for example, the operation of
addition is associative.  Second, the question of uniqueness has not
been examined.  It is possible there are other solutions satisfying
conditions (\ref{psiab}) and (\ref{comp}), and even that one could
satisfactorily replace the second of these conditions with a weaker (and
smoother)
requirement along the lines 
that $\hat{A}$ is ``close'' to $\hat{B}$ whenever ${\cal A}$ is
``close'' to ${\cal B}$.  Third, it might be possible to use the existence of
an algebra for POM observables to provide a ``cleaner'' formulation of
QM, not requiring any {\it a priori} distinction between POM
observables and Hermitian observables (which, for example, the usual mapping
between classical and quantum Hamiltonians requires).  For example, the
optical phase observable can now simply be 
defined {\it algebraically}, in direct analogy to the classical formula,
as the combination $e^{i\Phi} := N^{-1/2}a$.

Applications of the generalised statistical deviation and distance, discussed in section 4, to determining the optimal estimate of an observable from a given measurement and to finding general ``disturbance'' and ``joint-measurement'' uncertainty relations, have recently been given (since this paper was first prepared) \cite{ozawa,halljoint}.

\end{document}